\documentclass[referee]{aa} 
\usepackage{natbib}
\usepackage{psfig,epsfig,amssymb,alltt}
\usepackage{latexsym}
\usepackage{graphicx}
\graphicspath{{PS/}}
\usepackage{psfig}
\usepackage[english]{babel}
\usepackage{indentfirst}

\vspace{1cm}
\newcommand{\be}{\begin{eqnarray}}
\newcommand{\ee}{\end{eqnarray}}

\newcommand{\bitem}{\begin{itemize}}
\newcommand{\eitem}{\end{itemize}}

\begin{document}
\title{Sunyaev-Zel'dovich clusters reconstruction in multiband bolometer camera surveys}
       
\author{S. Pires\inst{1} \and J.B. Juin\inst{1} \and D. Yvon\inst{1} \and  Y. Moudden\inst{1} \and S. Anthoine\inst{2} \and E. Pierpaoli\inst{3}}

\institute{
CEA-CE Saclay, DSM/DAPNIA, F-91191 Gif-sur-Yvette Cedex, France
\and
PACM, Princeton University, Princeton, NJ, 08544 USA
\and
California Institute of Technology, Mail Code 130-33, Pasadena, CA, 91125 USA
}

\offprints{yvon@hep.saclay.cea.fr}
\mail{yvon@hep.saclay.cea.fr}
 
\date{\today}


\abstract{We present a new method for the reconstruction of Sunyaev-Zel'dovich (SZ) 
galaxy clusters in future SZ-survey experiments using multiband bolometer cameras such as Olimpo, APEX, or Planck. Our goal is to optimise SZ-Cluster extraction from our observed noisy maps. We wish to emphasize that none of the algorithms used in the detection chain is tuned on prior knowledge on the SZ -Cluster signal, or other astrophysical sources (Optical Spectrum, Noise Covariance Matrix, or covariance of SZ Cluster wavelet coefficients). First, a blind separation of 
the different astrophysical components which contribute to the observations is conducted
using an Independent Component Analysis (ICA) method. 
This is a new application of ICA to multichannel astrophysical data analysis. 
Then, a recent non linear filtering technique in the wavelet domain, based on multiscale entropy and the False Discovery Rate (FDR) method, is used 
to detect and reconstruct the galaxy clusters. Finally, we use the Source Extractor software to identify the detected clusters. The proposed method was applied on realistic simulations of observations that we produced as mixtures of synthetic maps of the four brightest light sources in the range 143~GHz to 600~GHz namely  the Sunyaev-Zel'dovich effect, the Cosmic Microwave Background (CMB) anisotropies, the extragalactic InfraRed point sources and the Galactic Dust Emission. We also implemented a simple model of optics and noise to account for instrumental effects. 
assuming nominal performance for the near future SZ-survey Olimpo, our detection chain recovers 25\% of the cluster of mass larger than $10^{14} M_{\odot}$, with 90\% purity.
Our results are compared with those obtained with the published algorithms of  \cite{algo:Pierpa05}.
As for global detection efficiency, this new method is impressive as it provides comparable results to \citet{algo:Pierpa05} in the high-purity/low completeness region, being however a \emph{blind} algorithm: (ie without using any prior on data to be extracted).  
 }

\maketitle 
\markboth{SZ Clusters Reconstruction}{}


\section{Introduction}
	In the last years, spectacular advances in cosmology happened through observation of the
	sky at millimeter and 
	sub-millimeter wavelengths. The Boomerang \citep{exp:Boomerang}, MAXIMA \citep{exp:MAXIMA}, Archeops \citep{exp:Archeops}, WMAP \citep{exp:WMAP}, CBI \citep{exp:CBI} and DASY \citep{exp:DASY} experiments measured the anisotropies of Cosmic Microwave Background with high precision. Combined with other observations, such as distant supernovae of \citep{exp:HighZ, exp:SNCP} and/or Clusters \citep{algo:WhiteCarls, phys:PSW01} or Large Scale Structure Observation \citep{phys:Eisenstein}, the former experiments allowed to place tight constraints on parameters of generic cosmological models.
	These results gave rise to new questions. Consensus cosmological models assume that the 
	large scale evolution of the Universe is driven by Dark Matter and Dark Energy. Dark Matter has been sought for 30 years, 
	and most candidates have been rejected in light of experimental results. Dark Energy will be one of the toughest challenges of modern cosmology. One way to learn about the nature of dark energy, is to study the evolution of the Universe at late times during which dark energy is thought to have impacted the formation of large scale structures. Cluster surveys are an ideal tool for this purpose.

\subsection{Cluster detection in SZ survey}  
	Large scale structures can be studied using several probes: optical galaxy catalogs, X-Ray cluster surveys such as XMM-LSS \citep{exp:XMM-LSS}, 
	SZ cluster surveys such as Olimpo \citep{exp:Olimpo}, APEX, AMI \citep{exp:AMI}, AMIBA \citep{exp:AMIBA} and weak shear such as CFHT-LS ``wide''. 
The mass and redshift distribution of clusters of galaxies: $dN/dzdMd\Omega$ (where M, z and $\Omega$ are the cluster mass and redshift, and the solid angle respectively), is predicted with fair precision using Halo models and N-body simulations. SZ cluster, X-Ray and Optical surveys measure quantities related to cluster mass through complex phenomena, such as interstellar gas heating processes, clustering, processes of stellar formation and evolution, that are still not fully understood. The SZ signal from galaxy clusters is believed to be the simplest direct observable of cluster mass, though some uncertainties on extragalactic gas heating mechanisms remain. Measuring the cluster mass function will require these systematics to be understood. Pioneering SZ cluster surveys will take their first data this year. The Planck satellite survey will follow shortly \citep{exp:Planck}. In this paper we prepare for their analysis, using simulated data.

JB. Melin \citep{phys:MelinThese} following \cite{algo:Herranz2002a, algo:Herranz2002b}  developped a cluster extraction tool based on an optimised filter. The filter is optimised to damp the noise as well as the other astrophysical components. SZ signal is small compared to the other astrophysical sources, rendering the task of filtering difficult. In addition a filter damps some of the signal at the same time as it rejects parasitics. We then developped the idea of a 3 steps method using as a first step a source separation algorithm. The source separation algorithm sorts the noisy SZ signal from the other astrophysical components and the filtering step is left with the easier task of improving SZ cluster signal versus noise. Finally a detection algorithm extracts the SZ cluster candidates.

In the 100 to 600~GHz range, the brightest components of the sky are the Cosmic Microwave Background (CMB), the Infrared Point Sources, the Galactic dust emission and, swamped in the previous ones, the  SZ clusters. It follows that the true sky map $X_{\nu}(\vartheta, \varphi)$, in a given optic band centered on $\nu$, can be modeled as a sum of distinct astrophysical radiations as in
\begin{equation} \label{SkyBandModel1}
	X_{\nu}(\vartheta, \varphi)  =  CMB_{\nu}(\vartheta, \varphi)  + IR_{\nu}(\vartheta, \varphi)  + Gal_{\nu}(\vartheta, \varphi)  + SZ_{\nu}(\vartheta, \varphi) 
\end{equation}
where $\vartheta, \varphi$ denote spatial or angular indexes on 2D or spherical maps. 
 The \emph{observed} sky map is the result of the convolution of the true sky with the optical beam of the telescope plus the unavoidable contribution of instrumental noise $N_{\nu}(\vartheta, \varphi)$. 
Multiband signal processing in astrophysical applications deals with exploiting correlations between data maps at different $\nu$,  to enhance the estimation of specific objects of interest such as the power spectrum of the CMB spatial fluctuations,  or the mass function of galaxy clusters. Considering the case where the radiative properties of the sources are completely isotropic in the sense that they do not depend on the direction of observation, the above model can be rewritten in the following factored form:
\begin{equation} \label{SkyBandModel2}
X_{\nu}(\vartheta, \varphi) =  \sum_{i} a_{\nu, i} S_{i}(\vartheta, \varphi)  \,\,+\,\, N_{\nu}(\vartheta, \varphi)
\end{equation}
where $S_{i}$ is the spatial template and $a_{\nu, i}$  the emission law of the $i\,{\textrm{th}}$ astrophysical component. Although this is a coarse approximation in the case of Infrared Point Sources as described in section~\ref{sect:irps}, it is mostly valid for the other three components.  With observations available in $m$ channels, assuming the beam varies only slightly as a function of $\nu$, equation~(\ref{SkyBandModel2}) can be written in matrix form : 
\begin{equation} \label{SkyBandModel3}
X(\vartheta, \varphi) = A\,\, S(\vartheta, \varphi)  \,\,+\,\, N(\vartheta, \varphi)
\end{equation}
where $X(\vartheta, \varphi)$ is a vector in $\mathbb{R}^m$, $A$ is an $m\times n$ matrix, n is the number of contributing astrophysical components, $S(\vartheta, \varphi)$ is now a vector in  $\mathbb{R}^n$ and $N(\vartheta, \varphi)$ in $\mathbb{R}^m$.  Equation~(\ref{SkyBandModel3}) expresses that the observations consist of linear mixtures of astrophysical components with different weights and additive noise. Assuming no prior knowledge of $A$,  recovering the particular component of interest, the SZ effect map of galaxy clusters, can be seen as a blind source separation problem which we approach in this paper using Independent Component Analysis (ICA).  Such methods have been previously  used quite successfully in the analysis of astrophysical data from present or future multichannel experiments such as WMAP or Planck~\citep{ica:Del2003,ica:Maino2002,ica:Kur2003} where the focus is on estimating the map of CMB anisotropies. We concentrate here on separating the SZ component which has very different statistical properties compared to CMB and thus requires the use of other ICA methods. Then, due to the additive instrumental noise, the source separation process has to be followed by a filtering technique.\\ 
	
We use an iterative filtering based on Bayesian methods to filter the noise \citep{wlens:starck05}.
This method uses a Multiscale Entropy prior which is only defined for non-significant wavelet coefficients selected by the False Discovery Rate (FDR) Method \citep{wlens:benjamini95}.

Then, we identify SZ clusters in filtered maps using the Source Extractor software SExtractor \citep{algo:Bertin} which turned out to conveniently provide source extraction along with adequate photometry and de-blending.
	
\subsection{Outline of the paper}
	In the following, we first describe our sky model. We use the Olimpo SZ-cluster survey \citep{exp:Olimpo} as an example of a future ambitious SZ survey. We model the Olimpo instrument and produce simulated observed maps. With minor parameter changes, these algorithms will also allow us to simulate observed local maps typical of other bolometer camera 
	SZ surveys (i.e Planck or APEX observations). Then we explain the data processing methods used to recover  the map of SZ-cluster signal and the ``observed" cluster catalog. We quantify the efficiency of our detection chain and compare it to a recently published method, thanks to cooperation of the authors.
This paper is at the interface of astrophysics and applied mathematics. We tried to write an article understandable for both communities.

\section{Simulations of contributing astrophysical components}
	The input of our simulation is a list of cosmological model parameter values. We use typically a $\Lambda CDM$ 
	cosmological model with  $\Omega_t=1$, $\Omega_m=0.3$, $\Omega_b=0.04$, $\Omega_{dm}=0.26$, $\Omega_\Lambda=0.7$, $h=0.7$, $\sigma_8=0.85$, $n_s= 1$, $\tau=0.1666$ and a seed for the random number generator. This simulation was fully written in C++, so that we can mass produce our simulations on a PC farm, when needed. 
	
\subsection{Cosmic Microwave Background anisotropies}
	Big Bang cosmology models assume that the Universe has been evolving from a hot and dense plasma. While it expands, its contents is cooling down. At some time, so called \emph{decoupling}, when the temperature of the Universe lowered below $h \nu = k_BT \approx 0.2$~ eV, the mean free path of photons \emph{suddenly} became infinite so that a large majority of them have been propagating freely through the Universe ever since. These photons are nowadays observed having a black-body spectrum at temperature 2.728K \citep{exp:WMAP}: the Cosmic Microwave Background (CMB). Small anisotropies in the CMB are observed and interpreted as resulting from temperature and density fluctuations in the primordial plasma at decoupling time.\\
Bolometer cameras measure incoming power variations, while  scaning the sky. Thus the main blackbody spectrum (monopole) won't be observed. In the following simulations we also assume that the CMB dipole (Doppler effect due to Earth mouvement) has been subtracted from the data.
	
	In order to simulate a map of CMB anisotropies, we first enter the cosmological parameter list in the cmbeasy code \citep{algo:cmbeasy} and compute a Power Spectrum in Spherical Harmonics up to multipole 5000, of the CMB fluctuations. Then using a small field approximation \citep{algo:WhiteCarls}, we generate a random spatially correlated Gaussian field of primordial anisotropies with the previous power spectrum.  Figure \ref{RawMaps} shows a typical simulated CMB anisotropy map in units of ${\mu}K$.

\begin{figure*}[hhhtp]
\vbox{
\centerline{
\hbox{
\hspace{0.2cm}
}}
\vspace{0.2cm}
\centerline{
\hbox{
\hspace{0.2cm}
}}
}
\caption{The 4 physical components of the sky included in our simulation: 
{\bf upper right} is a map of the CMB's anisotropies in unit of $\mu$K, 
{\bf upper left} is the SZ Cluster map, in unit of y Compton, 
{\bf lower right} is the IR point source map, convolved with a beam of 2 arcmin, in Jy at 350GHz, finally 
{\bf lower left} is the Galactic dust map in unit of MJy/st at 100 $\mu$m.}
\label{RawMaps}
\end{figure*}

\subsection{IR point sources}\label{sect:irps}
	In our frequency range, infrared point sources are expected to be a significant contamination. Borys et al. from the
	SCUBA experiment published a list of infrared point sources at 350GHz and a $logN/LogS$ law from this data. 
	We extend this law to lower brightness, and generate a catalog of IR point sources in our field of view. 
	We assume that their optical spectra follow a grey body law parameterised as:
\begin{equation} \label{GreyBody}	
	F(\nu) = \nu^{\alpha} * Planck( \nu, T_0). 
\end	{equation}
where $Planck$ stands for Planck's black body spectrum, $\nu$ stands for frequency, $T_0$ is the temperature of the 
black body and  $\alpha$ is by definition the spectral index. For each IR point source, the spectral index was randomly picked between 1.5 and 2, and $T_0$ was set to 30~K. Thus each IR point source has a different optical spectrum. Then the IR point source is placed on 
	our map as a single pixel. IR sources are positioned randomly on the map but  a 
	user chosen fraction of them can be positioned within clusters of galaxies. We get an IR point source map in units 
	of Jy at 350~GHz as shown on Figure~(\ref{RawMaps}).

\subsection{Galactic Dust Emission}
	We know since FIRAS and IRAS that the galactic interstellar dust emits in the far infrared. We wanted for 
	these simulations to generate truly random galactic dust maps. So, instead of using the maps published by Schlegel et al.,  we followed 
	Bouchet and Gispert (\citeyear{algo:Bouch99}) and assumed that Galactic Dust was described by  a power spectrum of spatial correlations in $C_l  \propto 1/l^3$. We then generated randomly a map with the same algorithms as used for the CMB map simulations. We assume a homogeneous optical spectrum over our map given by equation \ref{GreyBody} with $\alpha$ =2, and $T_0$= 20K. Finally, we normalise the map rms at its observed values at high galactic latitude:
		at $\lambda$ =100 $\mu$m, the rms flux is  observed at 1ÊMJ/st. We get a galactic Dust Emission Map, shown on figure \ref{RawMaps} in natural units of MJ/st at 100 $\mu$m.

\subsection{Sunyaev-Zel'dovich Clusters} \label{SimuSZ}
	Hot intracluster gas is observed as a plasma at temperature  2 to 20~keV. A small fraction of CMB photons traveling through clusters Compton-scatter on electron gas. Since the electron kinetic energy is much larger than the CMB photon energy, CMB photons statistically gain energy when diffused. This effect, known as the thermal Sunyaev-Zel'dovich effect (\citeyear{phys:SZEffect}), results in a small distortion in the CMB Black-Body spectrum, in the direction of Clusters of galaxies.
	The probability that a photon scatters while crossing a cluster is given by:
\begin{equation} \label{ySZ}
	y = \int_{los}^{}  \frac {kT_e} {m_ec^2} n_e \sigma_T dl =  \frac {k \sigma_T} {m_ec^2}  \int_{los}^{} T_e n_e  dl 
\end{equation}
The optical spectral dependance is given by:
\begin{equation} \label{SZSpec}
	j(\nu) = 2 \frac{(kT_0)^3}{(h_pc)^2}  \frac{x^4e^x}{(e^x-1)^2} \Big(\frac{x} {tanh(x/2)-4}\Big)
\end{equation}
In these expressions, $T_e, n_e, m_e$ refer to the electron temperature, density and mass, respectively; 
$\sigma_T = 6.65\times10^{-25} \textrm{cm}^2$ is the Thomson cross-section and
 $x=k\nu/kT_{CMB}$ is the dimensionless frequency of observation in terms of unperturbed CMB temperature T= 2.728K .\\

Simulation of SZ Cluster map requires modelling large scale structure formation.	 
For this we translated in C++ the ICosmo semi-analytic model \citep{phys:ICosmo} which, starting from a cosmological parameter set, allows us to compute  the cluster density versus mass and redshift: $dn/dz dM d\Omega (M, z)$. 
Several fits to data and physical assumptions are involved and implemented in this computation. We run C-ICosmo using the options of concordance $\Lambda$CDM model and the  mass function from Seth and Thormen \citep{phys:ShethMoTormen2001}. The virial radius of a cluster is computed assuming a spherical collapse \citep{phys:Peebles93, phys:Peacok96}. This is the radius where the average overdensity is $\Delta_c =1 + \delta = 200 $ times the critical density (chosen independent of redshift). 
Gas heating is modeled according to \citep{phys:PierpaoliMT}, with the additional parameter $T^*$. Then we  randomly generate a cluster catalog in our field of view, according to density profile keeping those clusters which have an integrated Compton parameter Y larger than $3.10^{-12}$ st (corresponding to a mass cut of $10^{14}$ Solar Mass). \\

 An ellipticity is attributed to each cluster according to experimental observations \citep{phys:CoorayEllipticity2000}. We assume that the cluster gas density follows an elliptical beta model. For the time being, clusters are positioned on the map randomly and the orientation of the main axis of each ellipse is also chosen randomly. We project to account for two point correlations in cluster positions in 
the near future. A typical SZ cluster map in natural unit of y Compton is shown on Figure~\ref{RawMaps}.

\begin{figure*}[htp]
\vbox{
\centerline{
\hbox{
\hspace{0.2cm}
}}
\vspace{0.2cm}
\centerline{
\hbox{
\hspace{0.2cm}
}}
}
\caption{Simulated maps in Olimpo's four frequency bands.
{\bf upper right} is the 147 GHz Band, {\bf upper left)} is the 217 GHz Band, {\bf lower right} is the 385 GHz Band: CMB anisotropies, IR point sources and Galactic Dust blend in this band {\bf lower left} 500 GHz: IR point sources and Galactic Dust are the dominant features at high frequencies. SZ cluster signal is dominated by other astrophysical sources at all frequencies.}
\label{NoisyMaps}
\end{figure*}

\section{Modeling observations}
\subsection{Instrumental effects}
	A detailed model of observations can only be developed after data acquisition. In the following, we choose to use a very simple observation model, nevertheless representative of an ongoing project, the Olimpo balloon experiment. The Olimpo camera will observe the sky at 4 frequencies, 143, 217, 385 and 600 GHz. The millimeter-wave filters' bandwiths will be close to 30 GHz and are assumed to be of tophat shape. The beams at the 4 frequencies are assumed to be symmetric, Gaussian and of 3, 2, 2 and 2 arcmin FHWM respectively. The optical efficiency (mirrors, filters), the photon noise and the bolometer sensitivities and noise contributions are summarized by an equivalent observed noise temperature ${\bf n_{eqT}}$ on the Sky in unit of $\mu K_{CMB}/Hz^{-1/2}$. Based on the BOOMERanG experiment, typical values in the four channels of Olimpo are expected at 150, 200, 500, and 5000~$\mu \textrm{K}_{\textrm{CMB}}/\textrm{Hz}^{-1/2}$  respectively.

\subsection{Noise model}
	The simplest and most optimistic model is to assume that the observed noise along bolometer timelines will be white, in which 
	case, after projection (or \emph{coaddition}, \citep{algo:MIRAGE}) on 2D maps, pixel noise can be very easly computed as: 
\begin{equation} \label{PixNois}
	nois_{pix} =  n_{eqT} * \sqrt{ N_{Bol}*t_{pix} }
\end{equation}
where $N_{Bol}$ stands for the number of bolometers working in a particular frequency band, and $t_{pix}$ is the observation length on this pixel. 
High resolution SZ bolometer surveys will cover their large survey adding up small patches of the sky. Noise map patterns are likely to be complex, and will only be known after data has been acquired.
We chose in this first paper to begin with a simple case of homogeneous white noise on the full field of view.

\begin{center}
\begin{tabular}{|c|c|c|c|c|}
\hline
Frequency [GHz] & 143 & 217 & 385 & 600 \\
\hline
\hline
FWHM [arcmin] & 3,5 & 2 & 2 & 2 \\
\hline
White noise level  [$ \mu K / \sqrt{Hz}$] & 150 & 200 & 500 & 5000 \\
\hline
\end{tabular}
\end{center}

\subsection{Mixing model}
	We now have all the tools to compute simulated observed sky maps. For each frequency band, we compute a conversion 
	factor from the maps in their natural units into the observed unit at the bolometer level namely $pW/m^2/st$. The exception
	 is the IR point source map where, because of the random spectral indexes of the point sources, a conversion factor is computed for each 
	 point source. We sum the four physical components into a ``true" sky map which is then convolved  with the experimental beam. Then we add the noise. We end up with one map per frequency band (figure \ref{NoisyMaps}).
	These maps would be what the analysis team would recover from the data, after pointing reconstruction, parasitics 
	removal, de-correlation of instrumental systematics in the data, and map-making: a big analysis work. In the following, 
	we explain a set of algorithms optimised for recovering SZ cluster signals in the observed maps. 

\begin{figure}[htp]
 \centerline{
\hbox{
}
}
\caption{SZ component map extracted by JADE from the four observed noisy maps. The SZ cluster signal, subdominant at all observed frequencies, now appears clearly. No obvious leftovers from other astrophysical sources are seen. Remaining noise is small, because we prefiltered data before JADE processing, and we simulated the nominal noise levels of an ambitious project: Olimpo.}
\label{JADEMaps}
\end{figure}
%
%
%

\section{Separation of Astrophysical components }
The simulated observations generated according to the method described in the previous section, consist of linear mixtures of the four main sources of diffuse radiation as expected in the four channels of the Olimpo instrument. Focusing on separating the SZ map,  we want to make the most of the data in the four frequency bands (centered on 143, 217, 385 and 600 GHz) and exploit the fact that each map is actually a different point of view on the same scene : multichannel data is about processing these maps in a coherent manner. Next is a quick overview of the general principles of ICA for component separation from multichannel data. Then, we give a more detailed description of JADE, the specific ICA algorithm we used. 

\subsection{ Independent component Analysis}
Blind Source Separation (BSS) is a problem that occurs in multi-dimensional data processing. 
The overall goal is to recover unobserved signals, images or \emph{sources}  $S$ from 
mixtures of these sources $X$ observed typically at the output of an array of $m$ sensors.
The simplest mixture model takes the form of equation \ref{SkyBandModel3}:
\begin{equation}\label{model0}
X = A \, S \,\,+\,\, N
\end{equation}
where $X$ and $S$ are random vectors of respective sizes $m \times 1$, $n \times 1$ and $A$ is an 
$m \times n$ matrix. 
The entries of $S$ are assumed to be independent random variables. 
Multiplying $S$ by $A$ linearly mixes the $n$ sources into $m$ observed processes.

Independent Component Analysis methods were developed to solve the BSS problem, \emph{i.e.}  given a batch of  $T$ observed samples of $X$, estimate the mixing matrix $A$ and reconstruct the corresponding $T$ samples of the source vector $S$, relying mostly on the  statistical independence of the source processes.  Note that with the above model, the independent sources can only be recovered up to a multiplication by a \emph{non-mixing} matrix \emph{i. e.} up to a permutation and a scaling of the entries of $S$. Although independence is a strong assumption, it is  in many cases physically  plausible. The point is that it goes beyond the simple second order decorrelation obtained  for instance using Principal Component Analysis (PCA) :  decorrelation is not enough to recover the source processes since any rotation of a white random vector remains a white random vector.\\

Algorithms for blind component separation and mixing
matrix estimation depend on the \emph{a priori} model used for the probability
distributions of the sources~\citep{ica:3easy} although rather coarse assumptions can be made \citep{ica:tutorial, ica:icabook}. 
 In a first set of techniques,
source separation is achieved in a noise-less setting, based on the 
non-Gaussianity of all but possibly one of the components. 
Most mainstream ICA techniques belong to this category : JADE~\citep{ica:JADE}, 
FastICA, Infomax~\citep{ica:icabook}. In a second set of blind techniques, the components 
are modeled as Gaussian processes and, 
in a given representation (Fourier, wavelet, etc.), separation requires that the sources 
have diverse, \emph{i.e.} non proportional, variance profiles. 
For instance, the Spectral Matching ICA method (SMICA) \citep{ica:Del2003,ica:mou2005}, 
considers in this sense the case of mixed stationary Gaussian components in a noisy context : 
moving to a Fourier representation, the idea is that colored 
components can be separated based on the diversity of their power spectra. 

In the case where the main component of interest is well modeled by a 
peaked distribution with long tails (\emph{e.g.} Laplace distribution) as is the case with SZ maps, methods from the first set are expected to yield better results. Next is a description of JADE, the non-gaussian ICA method we used.

\subsection{JADE} 
The Joint Approximate Diagonalization of Eigenmatrices method (JADE) assumes a linear mixture model as in~(\ref{model0}) where the independent sources $S$ are non-gaussian 
\emph{i.i.d.}\footnote{The letters  \emph{i.i.d.} stand for independently and identically distributed meaning that each of the entries of $X$ at a given position $t$ are independent of $X$ at any other position $t'$ and that the distribution of $X$ does not depend on position. } random processes with the additional assumption of a high signal to noise ratio (\emph{i.e.} $N \approx 0$).The mixing matrix is assumed to be square and invertible so that (de)mixing is actually just a change of basis.   Although the noise-less assumption may not be true in the problem at hand, the algorithm may still be applied and in fact, a proper change of representation can get us closer to such a setting, as discussed in the next section~\ref{sect:wJADE}. 

As mentioned above, second order statistics do not retain enough information for source separation in this context: finding a change of basis in which the data covariance matrix is diagonal will not in general enable to identify the independent sources properly. Nevertheless, decorrelation is \emph{half the job}~\citep{ica:tutorial} and one may seek the basis in which the data is represented by maximally independent processes among those bases in which the data is decorrelated. This leads to so-called orthogonal algorithms: after a proper whitening of the data by multiplication with the inverse of a square root of the covariance matrix of the data $W$, one is then seeking a rotation $R$ (which leaves things white) so that $\hat{ S}$ defined by
 \begin{equation}
\hat{ S} = W^{-1} \, Y =  W^{-1}\, R \, X_{\textrm{white}}  = W^{-1}\, R \, W \, X 
\end{equation}
 and $\hat{B} = \widehat{A^{-1}} =  W^{-1}\, R \, W$ are estimations of the sources and of the inverse of the  mixing matrix.\\

JADE is such an orthogonal ICA method and, like most mainstream ICA techniques, it exploits higher order statistics so as to achieve some sort of \emph{ non linear decorrelation}. Precisely, in the case of JADE,  statistical independence is assessed using fourth order cross cumulants defined by : 
\begin{eqnarray}  \nonumber	 
F_{ijkl} & = & \textrm{cum}( y_i, y_j, y_k, y_l )   \nonumber    \\
  & =& \mathcal{E} (y_i y_j y_k y_l) - \mathcal{E} (y_i y_j)\mathcal{E} (y_k y_l) -\mathcal{E} (y_iy_l)\mathcal{E} ( y_j y_k)-\mathcal{E} (y_iy_k)\mathcal{E} (y_j y_k) \\
\end{eqnarray}
where $\mathcal{E}$  stands for statistical expectation and the $y_i$'s are the entries of vector $Y$ modeled as random variables. Then, the correct change of  basis (\emph{i. e.} rotation) is found by somehow \emph{diagonalizing} the fourth order cumulant tensor. Indeed, if the $y_i$'s were independent, all the cumulants with at least two different indices would be zero. As a consequence of the independence assumption of the source processes $S$ and of the \emph{whiteness} of $Y$ for all rotations $R$,  the fourth order tensor $F$ is well structured: JADE was precisely devised to take advantage of the algebraic properties of $F$. JADE's objective function is given by 
\begin{eqnarray}  \nonumber	 
  \mathcal{J}_{\textrm{JADE}}( R ) & =&  \sum _{ij}   \sum_{k \ne l} \textrm{cum}(  y_i, y_j, y_k, y_l )^2  
\end{eqnarray}
which can be interpreted as a joint diagonalization criterion. Fast and robust algorithms are available for the minimization of $\mathcal{J}_{\textrm{JADE}}( R )$ with respect to $R$ based on Jacobi's method for matrix diagonalization~\citep{ica:pham2001}. More details on JADE can be found in~\citep{ica:JADE,ica:tutorial, ica:icabook}.

\subsection{JADE in wavelet space }\label{sect:wJADE}
We chose to use JADE after a Wavelet transform. Wavelets come into play as a sparsifying
\footnote{Data is sparse on a basis when this basis allows to describe that signal with a small number of coefficients. This is a higly desirable property, since noise is not expected to be sparse at the same time on such a basis. Choosing a sparsifying basis thus allows to enhance signal to noise ratio.} 
transform. Moving the data to a wavelet representation does not affect its information content and applying a wavelet transform on both sides of~(\ref{model0}) does not affect the mixing matrix and the model structure is preserved. However, the statistical distribution of the data coefficients in the new representation is different: wavelets are known to lead to sparse approximately \emph{i.i.d.} representations of structured data. Further, the \emph{local} (coefficient wise) signal to noise ratio depends on the choice of a representation. A wavelet transform tends to grab the informative coherence between pixels while averaging the noise contributions, thus enhancing structures in the data. 
Although the standard ICA model is for a noiseless setting, the derived methods can be applied to real data. Performance will depend on the detectability of significant coefficients \emph{i.e.} on the sparsity of the statistical distribution of the coefficients. Moving to a wavelet representation will often lead to more robustness to noise.    

Once the data has been transformed to a proper representation (\emph{e.g.} wavelets but also ridgelets and curvelets~\citep{starck:sta02_3} should the 2D or 3D data be strongly anisotropic), we apply the standard JADE method to the new multichannel  coefficients. Once the mixing matrix is estimated, the initial source maps are obtained using the adequate inverse transform (figure \ref{JADEMaps}).


\section{Clusters Restoration method - Noise filtering}
JADE has been used to separate the signals from four mixtures and four sources in the
 presence of Gaussian noise. We have added the expected experimental level of Gaussian noise 
 at each observed mixture map. In order to prevent bias induced by pixellisation along our simulation and detection algorithms, we overpixellised our observed maps, and thus pixel noise is enhanced. But all algorithms for BSS that require whitening are sensitive to additive noise. In order to minimise the impact of this noise at the ICA step, we chose to convolve our observed map before JADE with a gaussian of optical beams' width.
 Then in order to optimise signal recovery, we have to perform a filtering to remove remaining noise. We tested different methods for denoising. Thanks to the original simulated SZ map (without noise), we can easily compare the results of those filterings in the next part. 
 
\subsection{Linear Filtering}
\subsubsection{Gaussian filter}
A rather common linear filtering technique uses a Gaussian filter, 
generally isotropic. The standard method consists in convolving the 
observed map $S_{obs}$ with a Gaussian window G with standard deviation $\sigma_G$ :
\begin{eqnarray} {S}_G =  G * {S_{obs}}
\end{eqnarray}
\indent The filtering depends strongly on the value of $S_G$. This value is adjusted arbitrarly or based on a priori information.

\subsubsection{Wiener filter}
An alternative to Gaussian filtering is Wiener filtering which is a 
method that attempts to minimize the mean squared error between the
original and the restored signal. Wiener filtering consists in convolving the 
observed map $S_{obs}$ with a weight function that is to say by 
assigning the following weight to each mode in Fourier Space:

\begin{equation}
\hat{w}(k)=\frac{|\hat{S}(k)|^2}{|\hat{S}(k)|^2+|\hat{N}(k)|^2}
\end{equation}

where $|\hat{S}(k)|^2$ is a model of the map power
spectrum and is in practice derived from the data. 
The weight function makes it possible to attenuate or to remove part 
of the frequencies if the signal-to-noise ratio is low. 
The filtering depends on the model of the noise.
The Wiener filter is the optimal filter if both the signal and the 
noise are well modeled as Gaussian Random Fields. 
This condition is not verified for SZ maps
which display highly non-Gaussian features. Nevertheless, Wiener filtering 
generally outperforms the simple Gaussian filtering.

In the following paragraphs, we have tested a state of the art non linear 
filtering method in order to improve signal recovery.

\subsection{Multiscale Entropy method using False Discovery Rate (FDR)}
\subsubsection{Maximum Entropy Method (MEM)}
The Maximum Entropy Method (MEM) is commonly used in 
astronomy for image processing (see \cite{starck:sta01_1,wlens:marshall02,starck:book02}
for a full description). It is based on entropy and Bayesian methods. 
The Bayesian approach provides the means to incorporate prior knowledge 
in data analysis. Choosing the prior is one of the most critical aspects 
of Bayesian analysis. Here an entropic prior is used.
In data filtering, an entropy functional assesses the information content 
of a data set. Among several possible definitions of entropy, 
the most commonly used in image processing is the Gull and Skilling 
definition \citeyear{entropy:gull91}:
\begin{eqnarray}
 H_g(S) = \sum_{k}\sum_{l} \left[S(k,l) - m(k,l) - S(k,l) 
\ln\left(\frac{S(k,l)}{m(k,l)}\right)\right]
\end{eqnarray}
where $m$ is a model, chosen typically to be a sky background.
$H_g$ has a global maximum at $S=m$. However, MEM
does not allow negative values in the solution, which is a problem in 
experimental SZ data analysis, where we measure
fluctuations about zero. To overcome this problem, it has been
proposed \citep{entropy:hobson04} to replace $H_g$ by:
\begin{eqnarray}
\label{hobson}
 H_{+/-}(S) = \sum_{k} \sum_{l} \psi(k,l) - 2m - S(k,l) \ln\left(\frac{\psi(k,l) + S(k,l)}{2m}\right)
\end{eqnarray}
where $\psi(k,l) = \sqrt{S^2(k,l) + 4m^2}$. Here $m$ does not play the same role as in (\ref{hobson}). 
It is a constant fixed to the expected signal rms.

To overcome the difficulties encountered by the MEM to restore images 
containing both high and low frequencies, \cite{starck:pan96} have 
suggested a definition of entropy in a multiscale framework which we 
describe in the next section. It has been shown that the main drawbacks 
of the MEM (i.e. model dependent solution, oversmoothing of compact objects, \dots)
disappear.

\subsubsection{Multiscale Entropy}

\begin{enumerate}

\item{Multiscale Entropy definition:}

The Multiscale Entropy method is based on the standard MEM prior 
 derived from the wavelet decomposition 
of a signal. The idea	is to consider the entropy of a signal as the sum of 
the information at each scale of its wavelet transform. And the information of
a wavelet coefficient is related to the probability of its being due to noise.

The Undecimated Isotropic Wavelet Transform (UIWT) decomposes an $n \times n$ image $S$ as in :
\[ 
S(k,l) = {C_J}_{(k,l)} + \sum_{j=1}^{J} {w_j}_{(k.l)}, 
\]
where $C_{J}$ is a coarse or smooth version of the original image $S$ and $w_j$ represents the details in $S$ at scale $2^{-j}$ 
(see \citeauthor{starck:book02} for details). Thus, the algorithm outputs $J+1$ sub-band arrays of size $n \times n$. We will use an indexing convention such that $j = 1$ corresponds to the finest scale (high frequencies).

Denoting $H(S)$ the information relative to the signal and $h(w_{j}(k,l))$ the 
information relative to a single wavelet coefficient, 
the entropy is now defined as:
 \begin{eqnarray}
 \label{entropy} 
H(S) = h(C_J(k,l)) + \sum_{j=1}^l \sum_{k,l=1}^{N_j} h(w_{j}(k,l))
\end{eqnarray}
where $l$ is the number of scales and $N_j$ is the size of map in the band 
(scale) j.

\item{Entropy definition:}

The function $h$ in (\ref{entropy}) assesses the
amount of information carried by a specific wavelet coefficient.
Several functions have been proposed for $h$.
A discussion and comparison between different entropy definitions 
can be found in \cite{wlens:starck05}.
We choose the NOISE-MSE entropy \cite{starck:sta01_1} for the SZ
 reconstruction problem in which the entropy is derived
using a model of the noise contained in the data:
\begin{eqnarray}
\label{entropy1}
h(w_{j}(k,l)) =  \int_{0}^{\mid w_{j}(k,l) \mid } P_n(\mid w_{j}(k,l) \mid - u) 
   (\frac{\partial h(x)}{\partial x})_{x=u} du
\end{eqnarray}
where $P_n(w_{j}(k,l))$ is the probability that the coefficient
$w_{j}(k,l)$ can be due to the noise: $ P_n(w_{j}(k,l)) =
\mathrm{Prob}(W > \mid w_{j}(k,l) \mid) $.
For Gaussian noise, we have:
\begin{eqnarray}
 P_n(w_{j}(k,l)) & =  & \frac{2}{\sqrt{2 \pi} \sigma_j} 
 \int_{\mid w_{j}(k,l) \mid}^{+\infty} \exp(-W^2/2\sigma^2_j) dW \nonumber \\ 
 & = & \mbox{erfc}(\frac{\mid w_{j}(k,l) \mid }{\sqrt{2}\sigma_j})
\end{eqnarray}
and equation \ref{entropy1} becomes \ref{entropy2}
\begin{eqnarray}
\label{entropy2}
h(w_{j}(k,l)) = \frac{1}{\sigma_j^2} \int_{0}^{\mid w_{j}(k,l) \mid} u 
              \mbox{ erfc}(\frac{\mid w_{j}(k,l) \mid -u}{\sqrt{2} \sigma_j}) du 
\end{eqnarray}

The NOISE-MSE is very close to the $l_1$ norm
(i.e. absolute value of the wavelet coefficient) when the coefficient
value is large, which is known to produce good results for the
analysis of piecewise smooth images \citep{cur:elad03}.  

\item{Signal and noise information:}

The SZ component obtained by blind separation is swamped by noise.
The following algorithm assumes that the observed map can be decomposed as:
\begin{eqnarray}
S_{obs} = S + N
\end{eqnarray}
Then, we can decompose the information contained 
in our image in two components, the first one ($H_s$) corresponding to the non 
corrupted part, and the other one ($H_n$) describing a component which contains no 
information for us:
\begin{eqnarray}
H(S_{obs}(k,l)) = H_s(S_{obs}(k,l)) + H_n(S_{obs}(k,l))
\end{eqnarray}
For each wavelet coefficient $w_{j}(k,l)$, we have to estimate the fractions $h_n$
and $h_s$ of $h$:
\begin{eqnarray}
H(S_{obs}(k,l)) = \sum_{j=1}^l \sum_{k,l=1}^{N_j} h_s(w_{j}(k,l))+\sum_{j=1}^l \sum_{k,l=1}^{N_j} h_n(w_{j}(k,l)) \\
\end{eqnarray}

\end{enumerate}

\subsubsection{Multiscale Entropy Filtering} \label{MEFilter}
\begin{enumerate}
\item{Filtering:}\\
The problem of filtering the observed map $S_{obs}$ can be expressed as follows. 
We look for a filtered map $S_f$ such that the difference between $S_f$ 
and $S_{obs}$ minimizes the information due to the signal (to recover all 
the signal) and such that $S_f$ minimizes the information due to the noise 
(we want no noise). These two requirements are somehow competing. A tradeoff is necessary, because, on one hand, we want to remove all the noise (heavy filtering) and on the other hand, we want to recover the signal with fidelity. In practice, we minimize for each wavelet coefficient 
$w_{j}(k,l)$ :
\begin{eqnarray}
l(\tilde w_{j}(k,l)) = h_s(w_{j}(k,l) - \tilde w_{j}(k,l)) +\beta.h_n(\tilde w_{j}(k,l))
\end{eqnarray}
where $w_{j}(k,l)$ are the wavelet coefficients of the observed map $S_{obs}$, 
$\tilde w_{j}(k,l)$ the wavelet coefficients of the filtered map $S_f$ and 
$\beta$ is the so called regularization (trade-off) parameter.\\

\item{Selecting significant Wavelet coefficients:}\\
Whatever the filtering, the signal is always substantially modified. We want to fully reconstruct significant structures, without imposing strong regularization while eliminating efficiently the noise. The introduction of the multiresolution support \citep{starck:mur95_2}, helps to do so. The idea is to apply the previous regularization (i.e. filtering) only on the non-significant (noisy) wavelet coefficients \citep{starck:pan96}. The other components of the maps are left untouched.
The new Multiscale Entropy becomes:
\begin{eqnarray}
\tilde h(w_{j}(k,l)) =  {\bar M}(j,k,l)  h(w_{j}(k,l))  
\end{eqnarray}
where ${\bar M}(j,k,l) = 1 - M(j,k,l)$, and $M$ is the "multiresolution support" defined as:
\begin{eqnarray} 
M(j,k,l) = \left\{ \begin{array}{ll} \mbox{ 1 } & 
\mbox{ if
}   w_{j}(k,l) \mbox{ is significant} \\ \mbox{ 0 } & \mbox{ if }  
w_{j}(k,l)
\mbox{ is not significant} \end{array} \right. 
\end{eqnarray} 
M describes, in a Boolean way, whether the data contains information at a given scale $j$ and at a given position $(k,l)$. $w_{j}(k,l)$ is said to be significant if the probability that the wavelet coefficient is due to noise is small.  In the case of Gaussian noise, a coefficient $w_{j}(k,l)$ is significant if $\mid w_{j}(k,l) \mid > k \sigma_j$, where $\sigma_j$ is the noise standard deviation at scale $j$, and $k$ is a constant. Without an objective method for selecting the threshold, it is adjusted arbitrarly, generally taken between 3 and 5 \citep{starck:mur95_2}.  

\item{Selecting significant Wavelet coefficients using the FDR:}\\
The False Discovery Rate (FDR) is a new statistical procedure due to
\citep{wlens:benjamini95} which offers an effective way to select an adaptative
threshold to compute the multiresolution support . This technique has recently been described by 
\citep{wlens:miller01,wlens:hopkins02,wlens:starck05} with several examples of
astrophysical applications. The FDR procedure provides the means to adaptively 
control the fraction of false discoveries over total discoveries. 
The FDR is given by the ratio (\ref{fdr}), that is, the proportion of declared
active which are false positives:
\begin{eqnarray} 
\mathcal{FDR} = \frac{V_{ia}}{D_a}
\label{fdr}
\end{eqnarray}
where $V_{ia}$ is the number of pixels truly inactive declared active,
and $D_a$ is the number of pixels declared active.
The FDR formalism ensures that, {\it on average}, the False Discovery Rate is no larger than
$\alpha$ which lies between 0 and 1. This procedure guarantees control over the FDR
in the sense that:
\begin{eqnarray} 
\mathcal{E(FDR)} \leq \frac{T_i}{V}.\alpha \leq \alpha
\end{eqnarray}

The unknown factor $\frac{T_i}{V}$ is the proportion of truly inactive pixels.  A complete description of the FDR method can be found in \cite{wlens:miller01}.  In \cite{wlens:hopkins02} and \cite{wlens:starck05}, it has been shown that the FDR outperforms standard method for source detection.
In this application, we use  the FDR method in a multiresolution framework 
(see \citeauthor{wlens:starck05}, \citeyear{wlens:starck05}).
We select a detection threshold $\mathcal{T}_j$ for each scale.
A wavelet coefficient $w_{j}(k,l)$ is considered significant if its absolute value
is larger than $\mathcal{T}_j$ as seen below.

\item{Multiscale Entropy Filtering algorithm:}

Assuming Gaussian noise, the Multiscale Entropy restoration method
 reduces to finding the image $S_f$ that minimizes 
 $J({S_f})$, given the map $S_{obs}$ output of source separation with:
\begin{eqnarray}
J({S_f})= \frac{\parallel {S_{obs}}  - S_f \parallel ^2}
  {2\sigma_n^2} 
  + \beta \sum_{j=1}^{J} \sum_{k,l} \tilde h_n( ({\cal W} {S_f})_{j,k,l})    
\end{eqnarray}
where $\sigma_n$ is  the noise standard deviation in $S_{obs}$, $J$ is the number
of Wavelet scales, $\cal W$ is the Wavelet Transform operator and $\tilde h_n( w_{j,k,l})$ is the multiscale entropy only defined for non significant coefficient (outside de Support selected by the FDR thresholding). Full details of the minimization algorithm can be found in \cite{starck:sta01_1} 
The results presented in the next section, are obtained on a SZ map with a uniform Gaussian 
white noise but the method still holds for a non uniform Gaussian noise over the map. 

\end{enumerate}

\section{Clusters detection}
\subsection{Extraction algorithm - SExtractor}\label{SExParagraph}
We use a public source software for extraction of SZ-Cluster \emph{candidates} from the filtered maps: SExtractor \citep{algo:Bertin}. SExtractor turned out to be fast, convenient, and easy to configure. It handles conveniently Cluster Extraction, with observed sizes ranging from a few pixels up to angular diameters of a degree.
We used the  \emph{Noiseless} configuration, assuming that our filters are efficient enough for this assumption to be valid. We mainly use sources identification capability and the \emph{deblending} algorithms that turns out to be very usefull when extracting large-mass clusters. In order to recover as many candidates as possible, we set $\mbox{DEBLEND\_MINCONT}=0$ and $\mbox{DETECT\_MINAREA} = 5$ (the smallest cluster size after optical beam dilution is over $9$ pixels). Photometry of Cluster candidates is done using SExtractor's FLUX\_AUTO Mode. 
SExtractor photometry performs very well on our data:  the recovery error on integrated Compton flux Y is smaller than 4 percent, for Y larger than $2.5\,10^{-10}$\,sr (id $3\,10^{-3}$ arcmin$^2$).
Thus SExtractor outputs a catalog of Cluster candidates with their position, flux and size. This catalog is likely to be contaminated by false detections due to residual noise, or spurious point sources that have survived the Source Separation step. While processing observed data, we will have to live with this contamination. Using simulations, we quantify in the following the SZ-Cluster selection function (Completeness) and this contamination.

\subsection{Completeness and Purity of the recovered catalog}\label{ParagraphSourceExtractionReliability}
In order to identify in our catalog true SZ Clusters from contamination we used an association criterion: for each candidate cluster, we scan the generated cluster catalog (see \ref{SimuSZ}) for the closest cluster and store the angular distance between the two. Doing so, we observed a distribution with a strong peak at a distance smaller than 4 pixels (3.5 arcmin), corresponding to the true detected clusters, and a long flat tail corresponding to random associations. We then decided to tag candidates as true when their distance to the closest cluster in the simulated catalog is less than 4 pixels, neglecting the small fraction of random association passing this test. This criterion can and will be enhanced using an improved  distance involving recovered cluster flux and spatial distance.

\emph{Purity} is defined as the ratio of true detections to the total of detections:
\begin{equation}
\mbox{purity} = \frac{\mbox{true detections}}{\mbox{total detections}}
\end{equation}
And  \emph{Completeness} is defined as the ratio of true detections to the total number of clusters in the simulation:
\begin{equation}
\mbox{completeness} = \frac{\mbox{true detections}}{\mbox{total number of simulated clusters}}
\end{equation}
Completeness and purity are the too main performance criteria of a detection chain. The higher the better! We will use them to compare detection chains in the result section.

\section{Results}
In the following, we will quantify the performance of our algorithm chain in two steps. We will focus on the SZ map reconstruction: first the source separation step with JADE and then the additionnal FDR-MultiScale Entropy Method (ME--FDR) filtering step. The ME--FDR filter will be compared first to the classical Gaussian and Wiener filters. Then we will discuss assumptions and performances of our method relative to the one published in \citep{algo:Pierpa05}.

The comparison between two methods for astrophysical  image reconstruction can be performed in 
several ways. 
One may initially forget about the astrophysical nature of the image, and just investigate 
the characteristics of residuals. This technique is typically  used by mathematicians 
when dealing with picture of diverse nature.
An astrophysicist may be more inclined to  assess how well the objects of interest (SZ cluster of galaxies, in this case) are recovered: how many of them are found, and with what precision.
Finally, a cosmologist would want to know to which extent the reconstruction procedure limits his ability to estimate cosmological parameters. This implies consideration on the purity and completeness as functions of cluster mass threshold, as well as precision in the reconstruction of the (central or integrated) Compton parameter for such masses. 
In what follows, we will take the mathematician's point of view, followed by  a short astrophysicist's one: we will compare the map's reconstruction error.  Then we will focus on the SZ Cluster detection performance of the full detection chain, by comparing the recovered source catalogs to the Simulated SZ cluster catalog, in terms of purity and Completeness as defined in section  \ref{ParagraphSourceExtractionReliability}. This allows a parameter free quantification of performances and trade-offs involved in cluster detection, being mostly independant of photometry details. 
Note also that the cluster detection efficiency does not convey information on the spread of 
the input/output y parameter. A more detailed description of the selection function, contamination, a discussion on photometry issues and finally methods and of cosmological implications will be given in a following paper \citep{exp:JuinSZCosmo2005}.

\subsection{Real life with JADE}\label{RealJADE}

Figure \ref{JADEMaps} shows the recovered SZ map, after source separation. JADE performs very well on our maps: the SZ signal subdominant at all scales now appears clearly. No obvious leftovers from other astrophysical sources are seen. 
A few points are worth noticing. First JADE assumes and requires that input data have zero mean. This point is quite easy to meet in bolometer experiments since our bolometer cameras measure flux \emph{variations} while the telescope scans the sky.\\
Second, JADE looses calibration information while processing data. Sky components are separated, but calibration of each output map has to be restored. This fact might be considered as annoying, but it is of common observationnal procedure to calibrate a survey on the brightest sources in the field (which can be observed in follow up experiments). We chose in the following to calibrate our SZ maps using the 100 brightest clusters in the field. We average their recovered integrated flux (SExtractor FLUX\_AUTO mode), and scale the map to match the average integrated flux of the 100 brightest simulated Clusters. In the following, before computing statistical tests, all the maps have been normalised this way. We foresee to suppress this feature by taking into account prior knowledge (the optical spectral dependance of the SZ component) in future work.

Third, it is very important to minimise noise in the observed maps \emph{before} source separation (JADE) too. JADE was designed to run on noiseless data. Not surprisingly, it is quite sensitive to noise. If noise level is not minimised before JADE, then the recovered mixing matrix will be inaccurate, and the SZ Cluster map will be polluted by remains of the other astrophysical components, inducing efficiency loses. Once the mixing matrix has been carefully estimated, one can apply it to the unfiltered observed maps to extract the observed SZ cluster map, and then apply an optimised filter and cluster detection algorithms to extract the cluster catalog. We chose to apply, before JADE, to the four observed maps a simple Gaussian filter with the widths of the optical beams.

\begin{figure}[hhhtp]
\vbox{
\centerline{
\hbox{
\hspace{0.2cm}
}}
\vspace{0.3cm}
\centerline{
\hbox{
\hspace{0.2cm}
}}
}
\caption{Input SZ cluster map (Upper left)  map, and the 3 maps as recovered by Completeness and (upper right) Gaussian filtering $\sigma =2.5\,\rm arcmin^2$, bottom left, Wiener filtering and bottom right ME--FDR method. We chose to plot 25 $\rm deg^2$ maps, to point out differences that do not show on 400 $\rm deg^2$ maps}
\label{FilterMaps}
\end{figure}

\subsection{Filtering methods' performances}
Additional filtering is then applied to recover smaller clusters. We now quantify performances of the ME--FDR filter relative to the simpler filtering methods (Gaussian and Wiener). We chose to show results obtained with maps processed by JADE running in wavelet space, our best method. In a first stage, we will quantify the results by computing an error map, and its properties. Then we will compare the output of the extraction procedures using the three filters.

\subsubsection{SZ map reconstruction}
Figure \ref{FilterMaps} shows the maps after two classical filtering methods and our FDR multiscale entropy method. All maps look quite the same: in nominal noise condition of such ambitious experiments as future SZ Cluster survey, only statistical tests can sort between the 3 filtering methods. We computed the error map (filtered map, input map subtracted) for each filtered option. Computing the rms of the error maps divided by the rms of the simulated SZ map (dimensionless rms) leads to $\sigma_{Gauss}=0.617$, $\sigma_{Wiener}=0.602$ and $\sigma_{FDR}=0.570$ . 


%
%
%
\begin{figure}[htp]
\begin{center}
\end{center}
\caption{
Dimensionless standard deviation of error maps (filtered map, input SZ map subtracted) at each scale of the wavelet decomposition, for the three studied filters options. ME--FDR filtering method (red squares) is a significant improvement compared to simpler filtering methods (Gaussian, black dots and Wiener, blue triangles).}
\label{sigmal1}
\end{figure}

A more accurate test is to plot (see figure \ref{sigmal1}) the dimensionless standard deviation of error maps (filtered map, input map subtracted) at each scale of the wavelet decomposition: 
\begin{equation}
	\sigma_L = \frac{ \sigma_{Err, L} } { \sigma_{Sim,L} }
\end{equation}
 where $\sigma_L , \sigma_{Err, L}$ and $\sigma_{Sim,L}$, are the standard deviation of the error map and the input SZ map, selecting the scale $L$ of our wavelet transform. ME--FDR method performs better than simpler filters at all scales. Running our algorithms while changing the noise level showed that the higher the noise level, the larger the gain in using our ME--FDR filtering compared to simple Gaussian, or Wiener filters.

\begin{figure}[htp]
\begin{center}
\end{center}
\caption{Purity versus Completeness for the three filtering options presented in this paper. Completeness + ME--FDR (red squares) out performs the other filter options Gauss (black dots) and Wiener (blue triangles), allowing larger Completeness at any required level of purity. In the exemple presented here we simulated $9942$ clusters according to $\Lambda CDM$ distribution, with a threshold value on $Y_{Comp}$ of $2\,10^{-12}$\,st: note that the Completeness is normalized to this total number of simulated cluster. The insert is a zoom on the high purity region of the graph.}
\label{purecompl1}
\end{figure}

\subsubsection{Full extraction chain}
Then considering our goal of detecting clusters, the relevant test is to compare the recovered 
catalog to the simulated cluster catalog, input of the simulated map. Observed cluster catalogs are extracted and processed as explained in \ref{ParagraphSourceExtractionReliability}.  
In a future paper (JB. Juin et al.) we will present selection function, contamination, a discussion on photometry issues and finally constraints on cosmology.
In this paper where we discuss detection procedures the relevant information is the curve of purity versus Completeness while rising threshold on cluster observed flux (figure \ref{purecompl1}). As expected, the higher the threshold, the higher the purity, and the lower the Completeness. 
We see once more that ME--FDR out performs the other filter options, allowing larger Completeness at any required level of purity.

\subsection{Comparison with another Wavelet based methodology}
In what follows, we will compare the method described above with the one presented in 
 \citet{algo:Pierpa05}.
 In line with the spirit of the previous sections, we  will compare 
 the general performance of the two methods in reconstructing the SZ images and 
 then assess a global purity and completeness, as defined in section  \ref{ParagraphSourceExtractionReliability}.
As we  are  not assessing  here  the goodness of the $y$ parameter reconstruction,   a full comparison with the  \citet{algo:Pierpa05} results (i.e. purity
and completeness as a function of cluster mass threshold, spread of 
the input/output y parameter relation ) is not possible at this time.

We will first describe the \citet{algo:Pierpa05} method, and then preform the comparison 
according to the mentioned criteria.

\subsubsection{Description of the alternative image reconstruction method}
The method presented in \citet{algo:Pierpa05}, is formally a \emph{non-blind} component separation that has been optimized in order to recover galaxy clusters. 
 In this method  component separation and the deconvolution of the beam effect are done in one step by computing the Bayes Least Square estimate under a Gaussian scale mixture model of
"neighborhoods" of wavelet coefficients. These "neighborhoods" are sets of wavelet coefficients which are associated with the same location and behave in a coherent manner. The optical frequency dependency of each component, the beam size and the noise level for each observation are assumed to be known. The procedure relies on the possibility to discriminate between the different components (CMB, clusters, Infra-Red point sources and Galaxy Dust) by modeling the joint probability of these neighborhoods by Gaussian Scale Mixtures (see \citet{algo:Pierpa05} for details).
 A Gaussian scale mixture is a random vector $x=\sqrt{z}\,u$, where $u$ is a Gaussian vector independent of the scalar random variable $z$ -- allowed to be non--Gaussian. \\
The model is determined by the probability distribution $p_z$ and the covariance matrices of $u$. The covariance matrices have to be adjusted at each scale of the wavelet decomposition as a function of the resolution of the observed maps. They are computed prior to running the algorithm on simulated maps of
each components at the correct resolution. The probability distribution $p_z$ is also component dependent. \citet{algo:Pierpa05} investigate several different possibilities for $p_z$, showing 
that different distributions lead to very different results in the mass reconstruction.
For example, the Gaussian assumption ($Pier1$ in the following) is most adequate to recover the maximum number of clusters but is however inefficient in reconstructing the appropriate central intensity of these clusters; while a non-Gaussian distribution modeled on the simulated SZ map ($Pier2$) is most adequate to recover the input Comptonisation parameter for bright clusters while missing the reconstruction of very low intensity ones.

\begin{figure*}[htp]
\vbox{
\hbox{
\hspace{0.1 cm}
\hspace{0.1cm}
}}
\caption{Maps computed by two of the algorithms presented in the following of 25 $deg^2$ size.  Left is input SZ Cluster map. Centered is the map output of Completeness after ME--FDR filtering. Right is the map output of the Pierpaoli et al. method, using the Gaussian assumption for probability distribution.}
\label{FilterMapsPierpaoli}
\end{figure*}

In \citet{algo:Pierpa05} the focus is on reconstructing the signal of the most massive clusters, 
since those are the ones which are likely to lead to the most precise constraints on cosmological parameters.  To this aim, the distribution $Pier2$  is  preferable to  $Pier1$ .
 As this paper is more focused on cluster detection efficiency (also including low--intensity clusters) and disregards the precision of the reconstruction,  we will consider here the Gaussian prior for $p_z$ ($Pier1$ in what follows).
 We remind the reader that a Gaussian prior for $p_z$ reduces the estimator to a local Wiener filtering in 
 Wavelet space, as all information about the signal non--Gaussianity is lost.
The result of \citet{algo:Pierpa05} is a set of beam--deconvolved maps (one for each physical component considered) which can be directly 
compared with the input ones, or with other method's.
The cluster's $y$ map is the one we are  using here for  comparison. As for cluster detection, we will present results obtained with the code used in \citet{algo:Pierpa05}, as well as the SExtractor code adopted here.

\begin{figure}[htp]
 \centerline{
\hbox{
}
}
\caption{Standard deviation versus scale for SZ cluster maps recovered by Completeness ME--FDR method (red squares) and Pierpaoli et al. Pier1 method (blue triangles). The top axis reports the wavelet's typical scale (in arcmin) corresponding to the index  on the bottom axis.}
\label{sigmal2}
\end{figure}

\subsubsection{Map comparison}
In figure  \ref{FilterMapsPierpaoli} we present the maps that will be used for the comparison.
While both maps recover well the intense clusters, the Pier1 processed map shows more low-intensity structure than the ME--FDR map.
This could be due to a better reconstruction of low--intensity clusters, as well as undesirable noise.
 Computing the Pier1 error map rms, we find $\sigma_{Pier1} = 0.589$, which should be compared to 0.570 for the Completeness ME--FDR.
An analysis of the residuals'  normalised standard deviation  for different  wavelet scales (shown in  figure \ref{sigmal2}) shows that  the two methods mainly differ at angular scales equal and above 14 arcmin. The major contribution to the difference in the total rms is therefore not associated with the 
typical (low mass) cluster scale, but to much larger ones.

\begin{figure}[htp]
 \centerline{
\hbox{
}
}
\caption{Purity versus Completeness curves for the four software detection chains studied in the following. We use red squares for Completeness-FDR+SExtractor, black dots for Pier1 and blue triangles for Pier2 with peak detection, black diamonds for Pier1 using SExtractor for source detection. The inserted plot is a zoom on the high purity region of the graph. }
\label{purecompl2}
\end{figure}

\subsubsection{Detection efficiency}
We compare here the detection efficiency defined in section \ref{ParagraphSourceExtractionReliability}. 
For consistency with our detection chain (see section \ref{RealJADE}), we recalibrate 
the Pierpaoli's maps  so that the average integrated Compton Flux of the 100 brightest clusters match the value observed in the input SZ cluster map (small correction).
Figure \ref{purecompl2} presents the curves of Completeness versus purity for Completeness FDR-ME + SExtractor,  Pier1, Pier2 using the peak finding algorithm presented in \citet{algo:Pierpa05} to detect clusters and the Pier1 method using SExtractor (PierNew). 
As expected, Pier2 is not as good as the other methods at all purity, since the probability  $p_z$
used here is optimized for accurate recovering of the most massive cluster's central intensity
and not for cluster detection.
Pier1 (the Gaussian distribution) improves Pier2 results, especially when SExtractor is used to detect the
clusters (PierNew). 
At low threshold (low purity) PierNew recovers more clusters than ME--FDR.
The low--intensity structures visible in figure \ref{FilterMapsPierpaoli} contain
a sizable number a clusters. As Completeness FDR-ME is designed to filter out noise better, it also filters out some low--intensity clusters which then cannot contribute to the detection rate any longer.
At high purities, PierNew and Completeness FDR-ME provide equivalent performance, given the statistical uncertainly of the small number of clusters  involved in the high--threshold cut.

\section{Conclusion}
We simulated observed sky maps at frequencies of a typical large multiband bolometric SZ-Cluster survey. We implemented a complete software detection chain, working in 3 steps. First we use a source separation algorithm, that is based on a Wavelet transform and the JADE-ICA algorithm. Then, we filter the SZ-Cluster map using an FDR-Multiscale Entropy method. Finally, we detect the clusters on the filtered maps using the SExtractor software in a \emph{noiseless} configuration. This detection chain is  very efficient, yielding 25\% of clusters of mass larger than $10^{14} M_{\odot}$ detected with 90\% purity. 
We compare our detection algorithm to previously published wavelet based ones \citep{algo:Pierpa05}. 
We restrict our comparison criteria to a global detection efficiency, as defined in section \ref{ParagraphSourceExtractionReliability}. We find a detection efficiency in the high--purity region comparable to the Gaussian probability case (Pier1 method), which is the model that provides the best performance for this comparison among those presented in the \citep{algo:Pierpa05}.
 The ME--FDR detection efficiency slightly degrades at lower cluster intensity with respect to 
Pier1 method, as  ME--FDR filters out low--intensity clusters during the denoising procedure.
These results, however, are impressive as ME--FDR, unlike \citep{algo:Pierpa05} methods, is a \emph{blind} algorithm that makes no assumption on the physical properties of the signal to be recovered.

Our future algorithmic work will be focused on improving the source separation step. Physics brings, along with data, a lot of prior knowledge on sky components, and we will use a \emph{sensible} part of this knowledge, mainly optical spectrum knowledge, in the source separation algorithms. This will allow to overcome the intrinsic scaling indeterminacy of the blind linear mixture model and so prevent loosing track of map calibration as in section \ref{RealJADE}. Additionally, JADE assumes to run on noise-free data. We will design our source separation algorithm to handle noisy data. Of course detection efficiency doesn't provide all the information we want to know in order to do cosmology, as the accuracy of the reconstruction and photometry issues are also important. In our next papers \citep{exp:JuinSZCosmo2005} we will use these algorithms to compute selection functions, contaminations and constraints on cosmology foreseen for the upcomming bolometric SZ-cluster surveys.

\begin{acknowledgements}
The authors wish to thank A. R\'efr\'egier, R. Teyssier (CEA/SAp), JB. Melin and J. Bartlett (Univ.P7, APC), E. Bertin and JL. Starck (CEA/SEDI), for many private discussions.
E.P. is an ADVANCE fellow (NSF grant AST-0340648), also supported by NASA grant 
NAG5-11489.
\end{acknowledgements}
\bibliographystyle{aa}
\bibliography{filterSZ,ica,Simu}

\end{document}